\documentclass[12pt,preprint,dvips]{aastex}

\def \msun{$\mathrm{M}_\odot$}
\def \rsun{$\mathrm{R}_\odot$}

\def \kms{km~s$^{-1}$}
\def \pap{Paper I}

\author{Daniel~C. Kiminki\altaffilmark{1}, M.~Virginia McSwain\altaffilmark{2}, 
Henry~A. Kobulnicky\altaffilmark{1}}

\altaffiltext{1}{Dept. of Physics \& Astronomy, University of Wyoming,
Laramie, WY 82070} 
\altaffiltext{2}{Dept. of Physics, Lehigh University, 
Bethlehem, PA 18015}

\begin{document}

\title{New Massive Binaries in the Cygnus OB2 Association}

\begin{abstract} 
As part of an ongoing study to determine the distribution of orbital
parameters for massive binaries in the Cygnus OB2 Association, we
present the orbital solutions for two new single-lined spectroscopic
binaries, MT059 (O8V) \&\ MT258 (O8V), and one double-lined eclipsing
binary (Schulte~3). We also constrain the orbital elements of three
additional double-lined systems (MT252, MT720, MT771). Periods for all
systems range from 1.5--19 days and eccentricities range from
0--0.11. The six new OB binary systems bring the total number of
multiple systems within the core region of Cyg OB2 to 11. The current
sample does not show evidence for a ``twin-heavy'' binary distribution.
\end{abstract}


\section{Introduction}
The formation of massive stars is the subject of extensive debate
\citep{Krum05, Bonnell98, Larson01}.  Current formation models are
based, in part, on the accumulated statistics of massive star binary
systems, such as \citet{Evans06}, \citet{Merm95},
\citet{Gies87}, and \citet{Garmany80}.  The extent to which these
statistics can be generalized is limited, however, because some
studies only sample small numbers of binary systems while others
include large samples of systems from radically different environments
with varying ages.  A recent adaptive optics imaging study and
modeling analysis of Sco OB2 by \citet{Kouwenhoven07} has attempted to
circumvent the selection effects of previous smaller studies in order
to obtain the true binary fraction of this cluster. They find a binary
fraction for B and A-type stars between 70\% and 100\%.
\citet{Chip07} utilized Monte Carlo simulations applied to the data of
\citet[hereafter \pap]{Kiminki07} to find a true binary fraction of
$f\geq70$\% for O and B stars in Cyg OB2. Unfortunately, the true
binary fraction of massive stars in young clusters is poorly known,
but it is one of several preserved parameters providing information
about the formation of stellar systems.  Additional parameters
preserved from the formation of young binary/multiple systems are the
eccentricity, separation, period, and angular momentum (in the form of
rotational velocities) \citep{Larson01}. Understanding the initial
conditions that lead to the formation of massive stars is aided by
reliable statistics on a sample of binary systems sharing a common
formation history.  As a step toward understanding the characteristics
of massive binaries in Cygnus OB2, we present a study of six new
massive systems from the radial velocity survey of 150 OB stars begun
in \pap.  The OB stars are probable members of the Association and lie
within $\sim$15~pc (at $d=1.7$~kpc) of the cluster center surveyed by
\citet{MT91}. The long term goal of this study is to provide an
accurate distribution of binary orbital parameters for massive stars
with an assumed shared formation history.

\S~2 provides observational details of the new spectroscopic
datasets. \S~3 discusses the measurement of radial velocities, the
search for periods in the radial velocity data, and the determination
of orbital elements via radial velocity curve fitting. \S 4 discusses
the orbital solutions to the single-lined binaries (SB1s), MT059 \&\
MT258 \citep[notation in the format of][]{MT91}, and the
double-lined binary (SB2), Schulte~3 (Cyg OB2 no.~3). \S~5 constrains
the orbital elements of three additional SB2s, MT252 \&\ MT720, and
MT771. Finally, \S~6 summarizes the results of the survey to date,
including the total number of OB binaries uncovered in the Cyg~OB2
core region.

\section{Observations}
\pap\ details the observations of this survey through 2005 October. We
have obtained seven additional datasets with the Longslit spectrograph
on the Wyoming Infrared Observatory (WIRO) 2.3 m telescope and the
Hydra spectrograph on the WIYN\footnote{The WIYN Observatory is a
joint facility of the University of Wisconsin, Indiana University,
Yale University, and the National Optical Astronomy Observatory} 3.5 m
telescope. Table~\ref{obs.tab} lists the observing runs at each
facility and the corresponding spectral coverages and mean spectral
resolutions. Instrumental setups at each facility for the first six
runs were consistent with previous observations described in \pap.

Observations at WIYN took place over seven nights on 2006 September
8--11 and 2007 July 3,4, \&\ 5. We used the Hydra spectrograph with
the Red camera, 2\arcsec\ blue fibers, and the 1200 l mm$^{-1}$
grating in second order to obtain three 1200 s exposures in each of
three fiber configurations (1500--1800~s on the second run) yielding a
maximum signal-to-noise ratio (S/N) of 80:1 (2006) and 130:1 (2007)
for the brightest stars. The spectral coverage was 3800--4500~\AA\ at
a mean resolution of $R\sim$4500. Copper-Argon lamps were used between
each exposure to calibrate the spectra to an RMS of 0.03~\AA\ (2 \kms\
at 4500~\AA), and the typical resolution was 1.0~\AA\ FWHM at
3900~\AA\ and 0.82~\AA\ FWHM at 4400~\AA. Spectra were Doppler
corrected to the heliocentric frame and checked against the radial
velocity standards HD182572 (G8IV), HD187691 (F8V), and the minimally
varying, amply sampled star MT083 before comparison to previous
datasets. Both runs were plagued by intermittent clouds.

Observations using the WIRO-Longslit spectrograph with the
600~l~mm$^{-1}$ grating in second order took place over 12 nights from
five runs of varying length (2006 June 16--20, 2006 July 15--16 \&\
19, 2006 October 7, and 2007 June 28--30). Exposure times varied from
600~s to 1800~s depending on weather conditions to obtain a maximum
S/N of 150:1 for the brightest stars. The spectral coverage was
3900--5900~\AA. Copper-Argon lamp exposures were taken after each star
exposure to wavelength calibrate the spectra to an rms of 0.03~\AA\
(1.9~\kms at 4800~\AA), and the typical spectral resolution
was 2.9~\AA\ FWHM at 4100, 2.4~\AA\ FWHM at 4800,
and 2.8~\AA\ FWHM at 5700 (2006 October and 2007 June had a
more uniform resolution of $\sim$2.5~\AA\ FWHM across the
chip). Spectra were Doppler corrected to the heliocentric frame and
checked against the radial velocity standard HD187691 and the
minimally varying, amply sampled star MT083 before comparison to
previous datasets. The 2006 June and July data were affected by
intermittent clouds and an abnormally high dark current arising from a
camera cooling problem.

We also obtained 8 nights of spectra using the WIRO-Longslit
spectrograph with the 1800~l~mm$^{-1}$ grating in first order (2007
August 28--September 4) to examine the H$\alpha$ and \ion{He}{1}
absorption lines in suspected SB2s. Exposure times varied from 600~s
to 1800~s depending on weather conditions and yielded a maximum S/N of
200:1 for the brightest stars. The spectral coverage was
5550--6850~\AA\ and Copper-Argon lamp exposures were taken after each
star exposure to wavelength calibrate the spectra to an rms of
0.03~\AA\ (1.4~\kms\ at 6400~\AA). The typical spectral resolution was
1.5~\AA\ FWHM across the chip. Spectra were Doppler corrected to the
heliocentric frame and checked against the radial velocity standard
HD187691 and the minimally varying, amply sampled star MT083 before
comparison to previous datasets. All new datasets were reduced using
standard IRAF reduction routines as outlined in \pap.

\section{Data Analysis and Orbital Solutions}
We obtained radial velocities, $V_r$, for the new observations via the
IRAF cross-correlation task XCSAO in the RVSAO package \citep{xcsao},
using a model stellar atmosphere \citep[TLUSTY]{LHub2003} of the
appropriate effective temperature and gravity as discussed in \pap.
We used the method outlined in \citet{mcswain03} to determine the
orbital parameters of each system.  We obtained an estimate of the
orbital period through an IDL\footnote{The Interactive Data Language
(IDL) software is provided by ITT Visual Information Solutions.}
program written by A. W. Fullerton which makes use of the discrete
Fourier transform and CLEAN deconvolution algorithm of
\citet{Roberts87}. The strongest peaks in the power spectrum of each
star were examined by folding the data at the corresponding period and
inspecting the $V_r$ curve visually. Orbital elements were then
procured by using the best period as an initial estimate in the
nonlinear, least-squares curve fitting program of \citet{Morbey74}.
The best solutions were attained by manually varying the initial
guesses of key orbital parameters until a minimum in the standard
deviation ($rms$) was found. Weights for each point were assigned as
the inverse of the 1$\sigma$ cross-correlation velocity error, where
error within the XCSAO task is calculated using,

\begin{equation}
\sigma_v = \frac{3w}{8(1+r)}.
\end{equation}

\noindent In this equation $w$ is the FWHM of the correlation peak and
$r$ is the ratio of the correlation peak height to the amplitude of
antisymmetric noise \citep{xcsao}. 

We also included an additional $10$ \kms\ error added in quadrature to
the 1~$\sigma$ cross-correlation errors for all spectra from the 2006
June and 2006 July datasets.  A varying offset of up to $\sim$10~\kms\
was measured for the radial velocity standard HD187691 and the amply
sampled, minimally varying system MT083 during these runs. We have not
been able to ascertain the source of this velocity anomaly, and we did
not encounter this problem during other runs.

The complete list of orbital elements for each binary system appears
in Table~\ref{orbparms.tab}. Listed within the table are the period in
days (\textit{P}), eccentricity of the orbit (\textit{e}), longitude
of periastron in degrees (\textit{$\omega$}), systemic radial velocity
(\textit{$\gamma$}), epoch of periastron (\textit{$T_O$}), primary and
secondary semi-amplitudes (\textit{$K_1$ \&\ $K_2$}), adopted or
calculated minimum primary and secondary masses in solar masses
(\textit{$M_1$ \&\ $M_2$}), primary and secondary mass functions in
solar masses (\textit{f(m)$_1$ \&\ f(m)$_2$}), spectral
classifications from this survey (\textit{S.C.$_1$ \&\ S.C.$_2$}), the
minimum primary and secondary semi-major axes in solar radii
(\textit{$a_1$sin~$i$ \&\ $a_2$sin~$i$}), and finally, the rms of the
fits (\textit{rms$_1$ \&\ rms$_2$}).

\section{Orbital Solutions to MT059, MT258, and Schulte~3}
\subsection{The SB1 MT059}
\begin{figure}
\epsscale{1.0}
\plotone{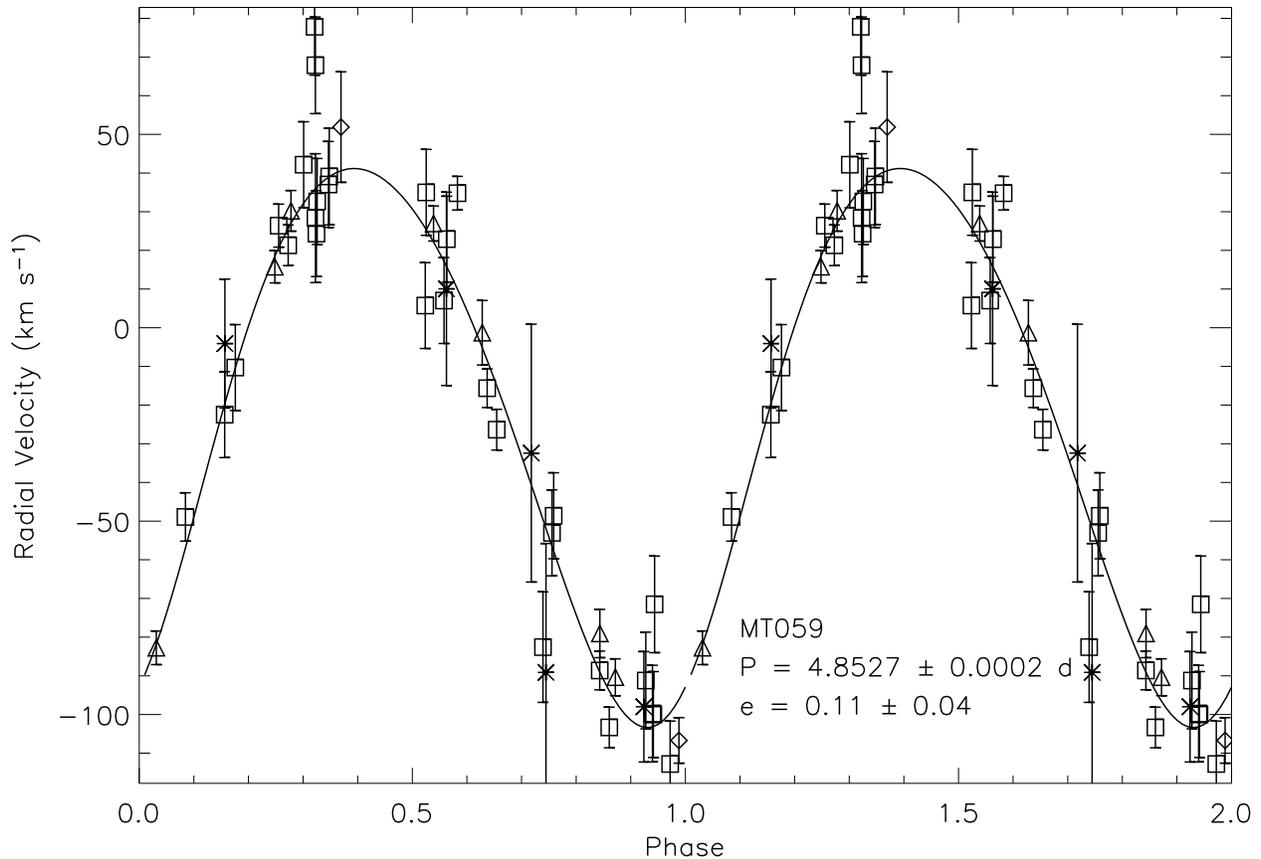}

\caption{$V_r$ curve and orbital solution to the O8V star, MT059. The
symbols correspond to the different observatories, where the stars are
observations taken with the Hamilton Spectrograph (Lick), the diamonds
are observations taken with HIRES (Keck), the triangles are
observations taken with Hydra (WIYN), and the squares are observations
taken with WIROspec or the Longslit Spectrograph (WIRO).
\label{MT059curve}}
\end{figure}

We have accumulated 45 observations of the O8V star, MT059 (Cyg OB2
No.~1), including six at Lick, two at Keck\footnote{The W.~M.~Keck
Observatory is operated as a scientific partnership among the
California Institute of Technology, the University of California, and
the National Aeronautics and Space Administration. The observatory was
made possible by the generous financial support of the W.~M.~Keck
Foundation.}, eight at WIYN, and 29 at WIRO. A total of 44 spectra
were included in the period search after eliminating one low S/N Lick
observation. The singular peak in the CLEANed power spectrum occurs at
a period of $P=4.852$ days. The power spectrum also contained a small
amount of low power noise but no potential
aliases. Figure~\ref{MT059curve} shows the $V_r$ curve and
best-fitting orbital solution with a period of $P=4.8527\pm0.0002$
days (the three most discrepant points in the figure belong to the
2006 June \&\ July runs and were given less weight to improve the
fit). Given this period, an assumed primary mass of $M_1=21.4\pm0.6$
\msun\ \citep{FM05}, and a semi-amplitude of $K_1=72.3\pm2.3$~\kms,
the solution yields a minimum secondary mass of
$M_{2}=5.1\pm0.3$~\msun\ (where the mass error is derived from the
error in the mass function).  This corresponds to a mid B star or
earlier \citep[interpolated from][]{drilling} and a mass ratio
$q=M_2/M_1\gtrsim0.24\pm0.02$. In the best spectra having S/N of 150:1
(WIRO-Longslit), we should be able to detect a secondary star's
spectral features at velocity separations of $(1+1/q)K_1$ if the
component luminosity ratio is larger than $L_2/L_1\simeq0.2$
(luminosity ratios have been chosen based on a set of spectra
combinations at various luminosity ratios and line separations
according to spectral type and theoretical mass ratios).  The absence
of spectral signatures from the secondary limits the luminosity ratio
to $L_2/L_1\lesssim0.2$, which, in turn, implies a secondary later
than B1V and an upper limit on the mass of $M_2\lesssim13.8\pm0.6$
\msun\ \citep[interpolated from][]{drilling}. Thus, the secondary mass
is constrained to $5.1 \lesssim M_2 \lesssim13.8 $~\msun\ and the mass
ratio is $0.24\pm0.02\lesssim q \lesssim 0.64\pm0.03$.  The upper
limit on the secondary mass also requires $i\gtrsim26^\circ$, however
this is a loose requirement given the uncertainty in O star masses
\citep{Massey05}. Additionally, the solution yields a minimum primary
semi-major axis of $a_{1}$sin~$i=6.9\pm0.2$ \rsun, an eccentricity of
$e=0.11\pm0.04$, a longitude of periastron of $\omega=212\pm17^\circ$,
and a residual $rms=11.0$~\kms. The residuals show no indication of
periodicity and have a Gaussian distribution.

The ephemerides for MT059 and MT258 are listed in Table~\ref{OC} and
include the star designation, date (\textit{$HJD-2,400,000$}), phase
(\textit{$\phi$}), measured radial velocity (\textit{$V_r$ }),
cross-correlation $1\sigma$ error (\textit{$1\sigma$~err}), and the
observed minus calculated velocity (\textit{$O-C$}).



\subsection{The SB1 MT258}
\begin{figure}
\epsscale{1.0}
\plotone{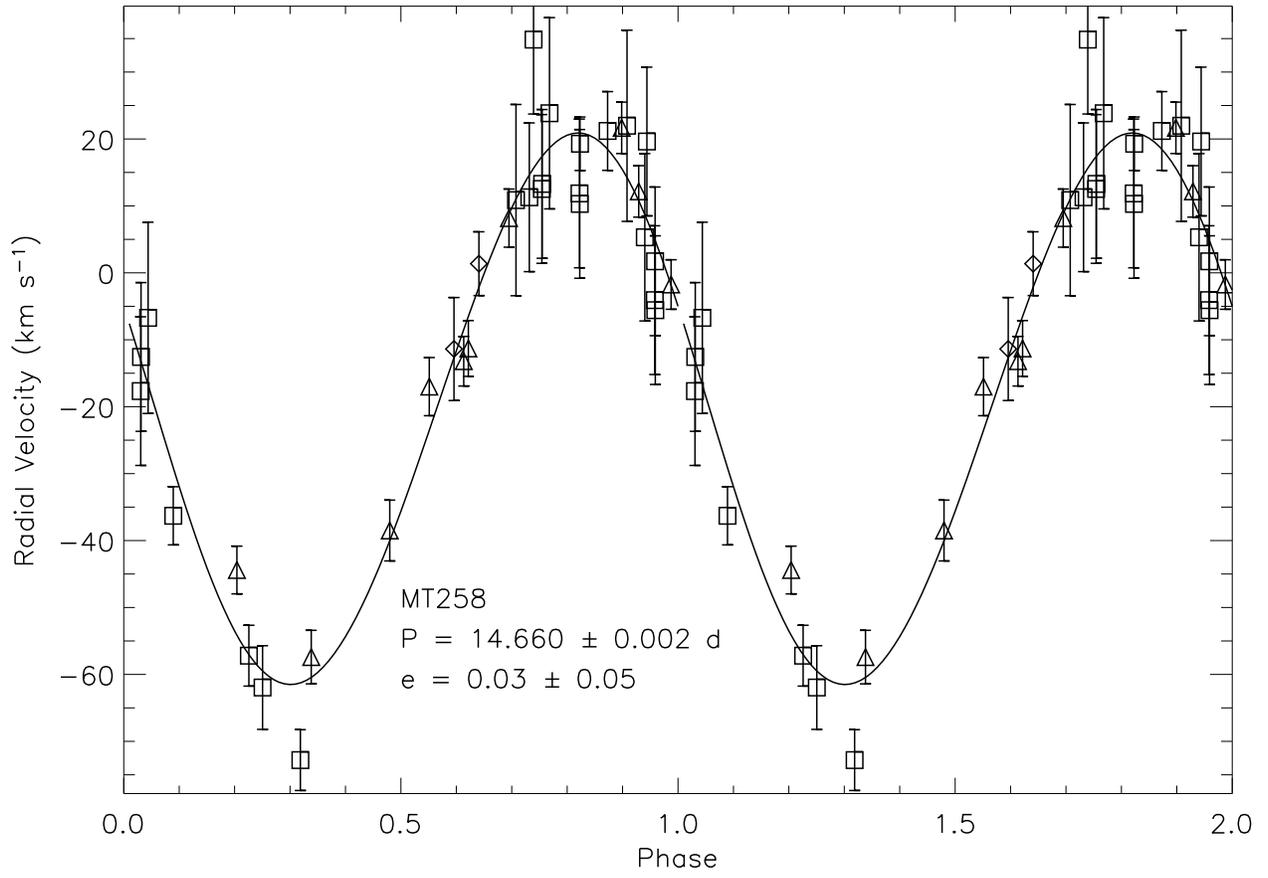}
\caption{$V_r$ curve and orbital solution to the O8V star,
MT258 using the same format as Figure~\ref{MT059curve}.
\label{MT258curve}}
\end{figure}

We observed another O8V star, MT258, 42 times, including two at Keck,
six at Lick, 13 at WIYN, and 21 at WIRO. The six spectra obtained at
Lick and one spectrum obtained at WIYN yielded low S/N and unreliable
velocities with large uncertainties and were subsequently removed. The
strongest signal in the CLEANed power spectrum for MT258 corresponds
to a period of $P=14.66$~days. In addition to some low power noise,
the power spectrum also contains an additional relevant signal
corresponding to a period of $0.49$~days. Insufficient variations over
the course of a night coupled with a visual inspection of the folded
$V_r$ curve reveals this period is an alias. The $V_r$ curve and best
fitting orbital solution shown in Figure~\ref{MT258curve}
($rms=6.7$~\kms) corresponds to a period of $P=14.660\pm0.002$~days
and an orbit with an eccentricity of $e=0.03\pm0.05$. Additionally,
the orbital solution provides a semi-amplitude of
$K_1=41.2\pm1.7$~\kms, a primary semi-major axis of
$a_{1}$sin~$i=11.9\pm0.5$~\rsun, a longitude of periastron of
$\omega=68\pm71^\circ$, and a minimum secondary mass of
$M_2\gtrsim4.1\pm0.2$~\msun, assuming a $21.4\pm0.6$~\msun\ primary
\citep{FM05}.  $M_2\gtrsim4.1\pm0.2$~\msun\ corresponds to a
late-to-mid B star \citep[interpolated from][]{drilling} and a mass
ratio $q\geq0.19\pm0.01$.  Given the comparable S/N between MT258 and
MT059 spectra, we should be able to detect a secondary star's spectral
features at velocity separations of $(1+1/q)K_1$ if the component
luminosity ratio is larger than $L_2/L_1\simeq0.4$.  The absence of
spectral signatures from the secondary and the semi-amplitude of
$K_1=41.2\pm1.7$~\kms\ limits the luminosity ratio to
$L_2/L_1\lesssim0.4$, which, in turn, implies a secondary later than
B0V and an upper limit on the mass of $M_2\lesssim17.5\pm1.0$ \msun\
\citep[interpolated from][]{drilling}. Thus, the secondary mass is
constrained to $4.1\pm0.2 \lesssim M_2 \lesssim17.5\pm1.0 $~\msun\ and
the mass ratio is $0.19\pm0.01\lesssim q \lesssim 0.84\pm0.05$.  The
upper limit on the secondary mass also loosely requires
$i\gtrsim18^\circ$.

\subsection{The SB2 Schulte~3}
Schulte~3 is listed as a probable binary system on the basis of
$H\alpha$ variability by \citet{Harries02}, and we report here the
double-lined nature of this system for the first time.  It is a
probable member of Cyg OB2, but it lies on the outskirts of the
cluster core surveyed by \citet{MT91}.  The spectral type of this
star, O8.5IIIf \citep[O9: in the Simbad database]{MT91} is not
entirely certain, and therefore the adopted absolute magnitude from
\citet{FM05} is also less certain. The spectrum does carry the same
diffuse interstellar band absorption (at 4428~\AA\ and 4501~\AA)
characteristic of known Cyg~OB2 stars \citep{Hanson03}. Preliminary
results of a concurrent photometric survey of this cluster show this
binary is also an eclipsing system with an estimated period of 4.75
days \citep[in prep]{Karen}.

\begin{figure}
\epsscale{1.0}
\plotone{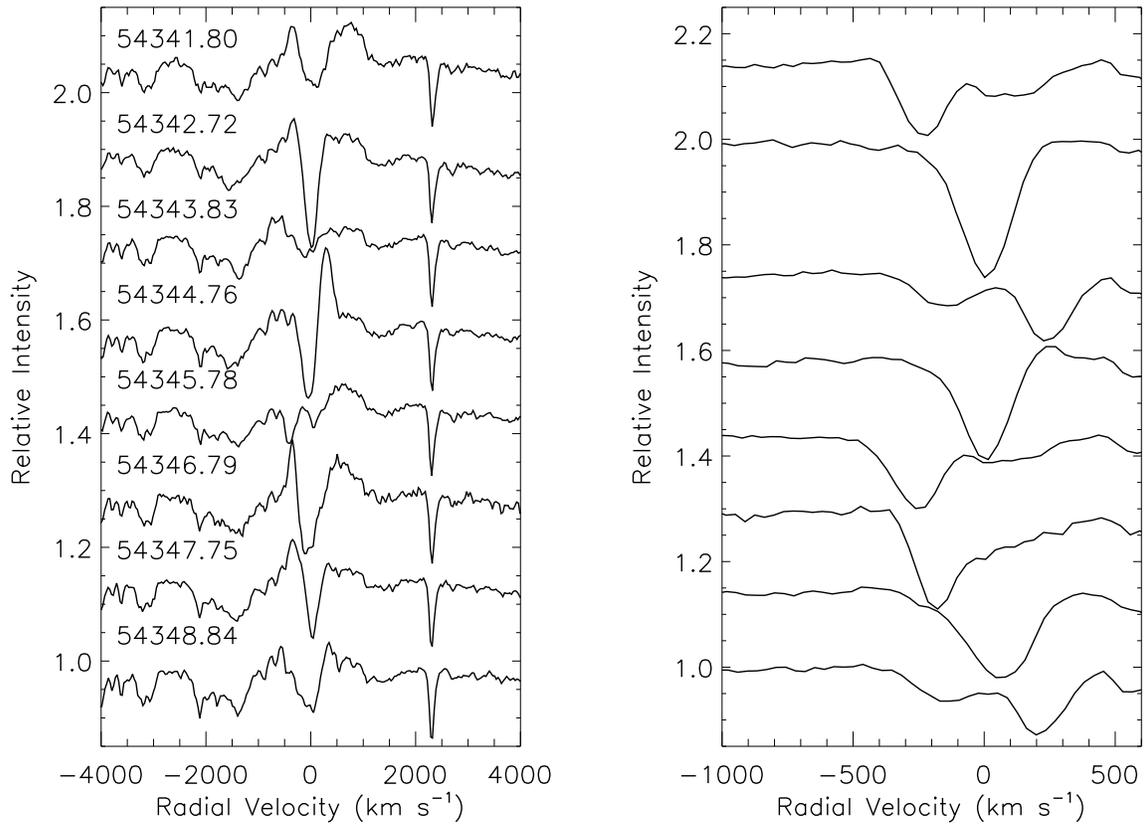}
\caption{H$\alpha$ $\lambda$6562.80~\AA\ (left) and \ion{He}{1}
$\lambda$5875.75~\AA\ (right) in velocity space for Schulte~3
between 2007 August 28 and 2007 September 4. Dates are
listed as HJD - 2,400,000 and apply to both left and
right panels. 
 \label{S3velprog}}
\end{figure}

\begin{figure}
\epsscale{1.0}
\plotone{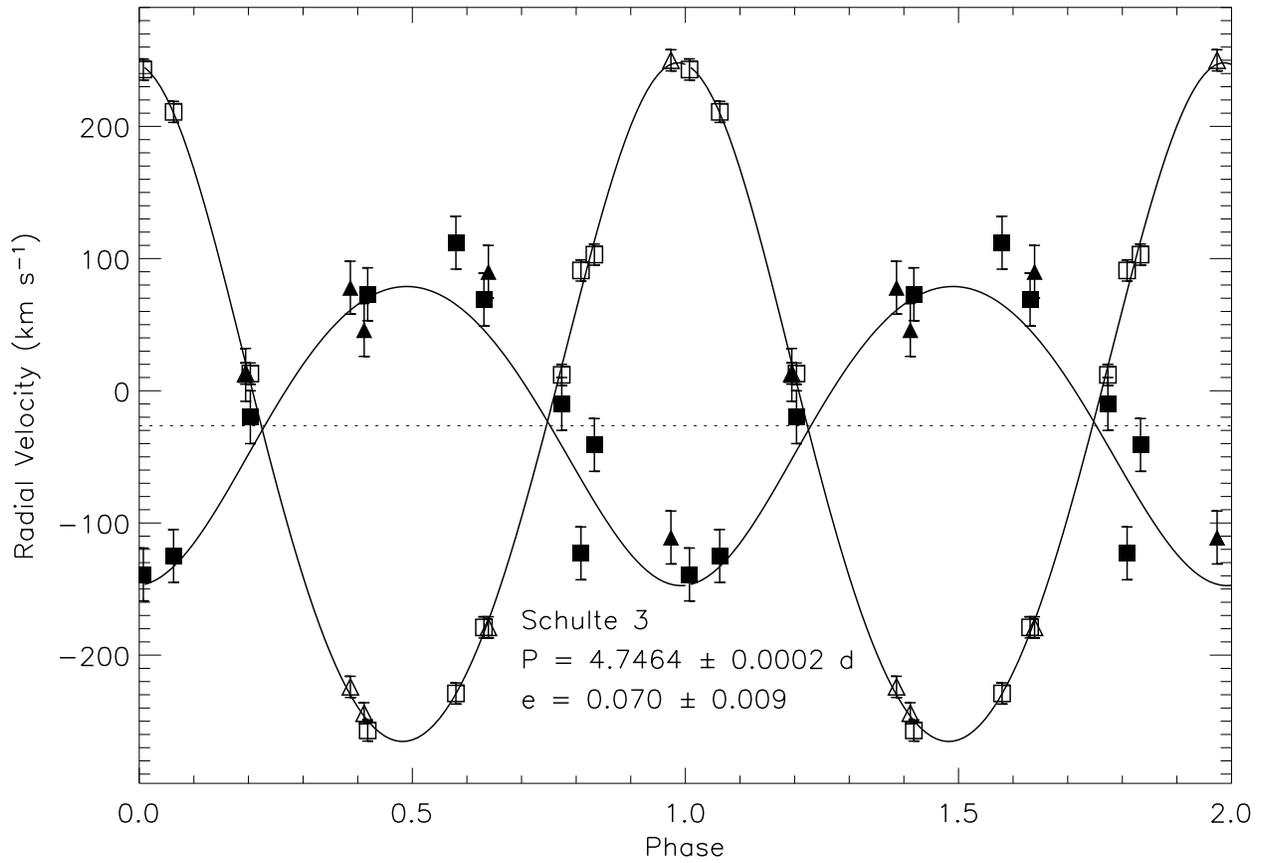}
\caption{$V_r$ curve and orbital solution for Schulte~3 using 14 of
the 18 observations. The filled
points correspond to the primary (O6IV:) and the unfilled points correspond to
the secondary (O9III).  The symbols represent the same observation
locations as Figure~\ref{MT059curve}.
 \label{S3fit}}
\end{figure}

We obtained 9 exposures of this star between 2000--2007 with Hydra at
WIYN, one exposure on 2000 September 19 at Keck, and eight exposures
between 2007 August 28 and September 4 at WIRO with the Longslit
spectrograph. The five exposures obtained at WIYN from 2006 September
8--11 show a progression from deblended, double-lined profiles to a
single, slightly asymmetric profile and back. Figure~\ref{S3velprog}
shows this same deblended/blended line progression in the 2007
August/September WIRO data. Using 14 of the 18 observations obtained,
we deblended the \ion{He}{1} $\lambda\lambda$4471, 5876~\AA\ lines by
fitting simultaneous Gaussian profiles with the SPLOT routine in IRAF
to obtain rough $V_r$ measurements. We repeated this method 10 times
while varying the baseline region used to define the continuum each
time. We then used the mean of each component's list of Gaussian
centers to compute the velocity and adopted the rms of this list as
our one sigma uncertainty. This results in an average uncertainty of
$\leq8$~\kms\ for the secondary and $\sim20$~\kms\ for the primary. We
observe a maximum line separation of $\sim$370~\kms\ for the
\ion{He}{1} lines. Emission in H$\alpha$ associated with the primary
makes it difficult to compare helium with hydrogen velocities, but the
maximum measured $V_r$ separation for \ion{He}{1} is consistent
throughout the data. Based on a visual inspection of the spectra in
Figure~\ref{S3velprog}, a period of 4.63~days from the CLEANed power
spectrum, and the preliminary results of \citet[in prep]{Karen}, we
estimated a period of 4.7~days. It should be noted however, that the
CLEANed power spectrum also contains a stronger signal corresponding
to a period of $1.29$~days, a $1-\nu$ alias (where $\nu$ is the signal
frequency in $days^{-1}$). However, a period of $1.29$~days does not
produce a sinusoidal folded $V_r$ curve for the primary or agree with
the other period estimates. We attribute the power of this signal to
the considerably smaller number of data points used in the period
search. The best orbital solution yields a period of
$P=4.7464\pm0.0002$~days for the secondary. The period, eccentricity,
epoch of periastron, and systemic velocity were applied as fixed
parameters to obtain the solution for the primary.  The residuals and
errors are considerably higher for the primary and likely a result of
emission in the helium line cores during a portion of the system's
orbit. The $V_r$ curve and orbital solution for Schulte~3 is shown in
Figure~\ref{S3fit}. The filled symbols correspond to the primary $V_r$
measurements and the unfilled symbols correspond to the secondary
$V_r$ measurements. With a low eccentricity of $e=0.070\pm0.009$ and
semi-amplitudes of $K_2=256.7\pm2.4$~\kms\ and
$K_1=113.2\pm14.5$~\kms, the solutions yield a nearly circular orbit
and a mass ratio of $q=0.44\pm0.08$. The calculated semi-major axes
are $a_{1}$sin~$i=7.4\pm0.9$~\rsun\ and
$a_{2}$sin~$i=16.7\pm0.2$~\rsun\ (i.e., a separation of
$a$sin~$i=24\pm0.9$~\rsun), and the calculated lower limit masses are
$M_{1}$sin$^3i=17.2\pm0.3$~\msun\ and
$M_{2}$sin$^3i=7.6\pm1.4$~\msun. The ephemeris for Schulte~3 in
Table~\ref{OC2} lists the star designation, date
(\textit{HJD$-$2,400,000}), phase (\textit{$\phi$}), measured radial
velocities (\textit{$V_1$~\&~$V_2$}), and the observed minus
calculated velocities (\textit{$O_1-C_1$~\&~$O_2-C_2$}). Errors
estimates are shown in parentheses.

\begin{figure}
\epsscale{0.9}
\plotone{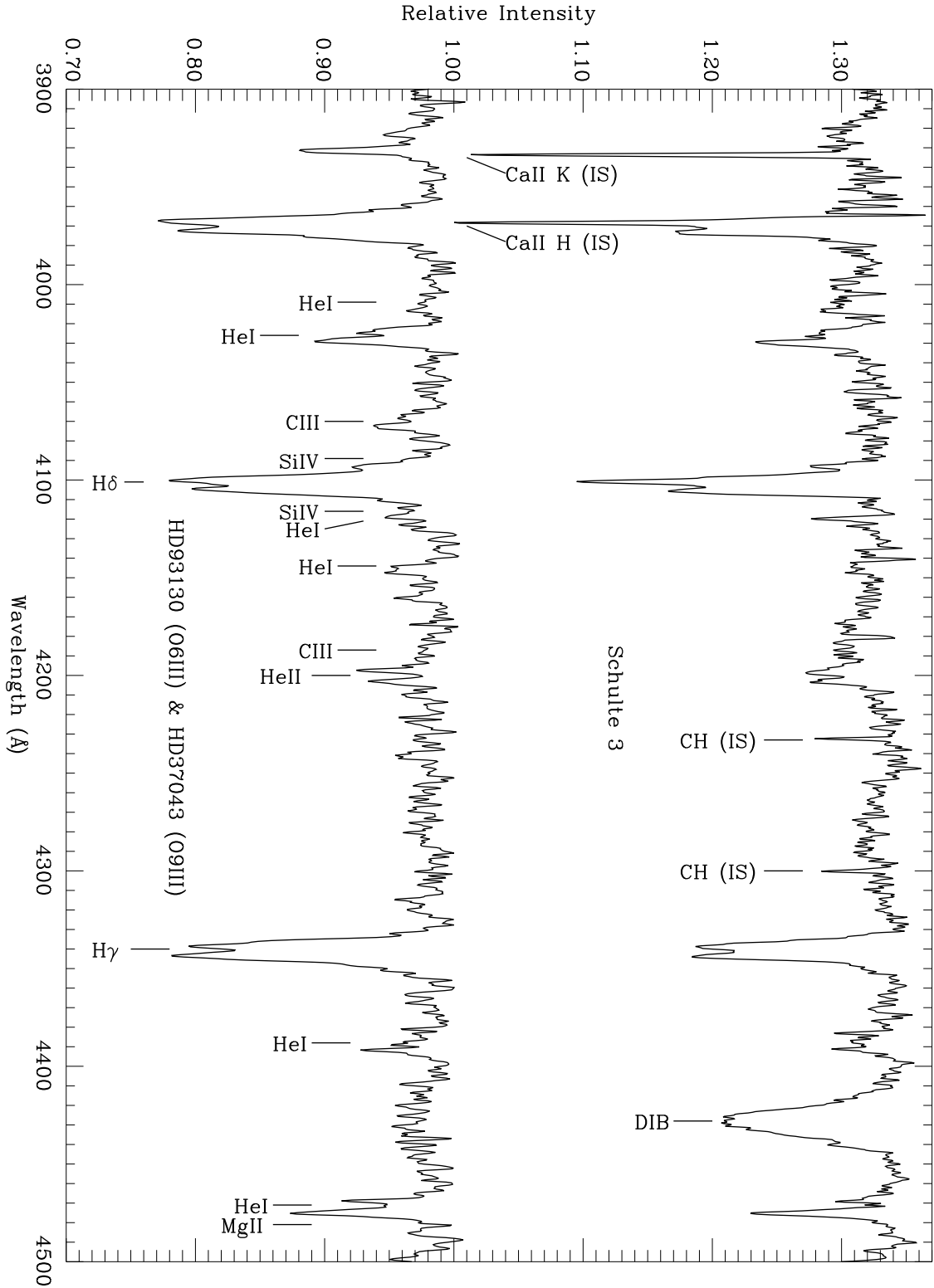}
\caption{A spectrum of Schulte~3 obtained on 2007 July 5 with WIYN
(top), and a composite of two shifted spectra from the \citet{WF90}
digital atlas (bottom) that best match the spectral types of the
primary and secondary of Schulte~3.  The primary is shifted -110~\kms\
and the secondary is shifted 257~\kms. Absorption features are labeled
at their respective intrinsic values.
 \label{S3comp}}
\end{figure}

The top spectrum displayed in Figure~\ref{S3comp} was obtained 2007
July 5 at WIYN. In this spectrum, the secondary's spectral features
(redshifted) are consistent with an O9III spectral type. Line ratios
were estimated by a combination of equivalent width and line depth
comparisons. The temperature-sensitive ratio,
\ion{He}{1}/\ion{He}{2}~$\lambda$4026:\ion{He}{2} $\lambda$4200~\AA\
($\sim$2:1), in addition to the presence of \ion{He}{1}
$\lambda$4009~\AA, the ratio of \ion{He}{1} $\lambda$4144:\ion{He}{2}
$\lambda$4200~\AA\ (1:2), and the ratio of \ion{He}{1}
$\lambda$4471:H$\gamma$ \&\ H$\delta$ (2:3) indicates the temperature
class is most likely an O9 \citep{WF90}. The luminosity indicative
ratio \ion{He}{1} $\lambda$4026:\ion{Si}{4} $\lambda$4089~\AA\ is
$\sim$2:1 and corresponds to a luminosity class of III
\citep{WF90}. The primary's features (blueshifted) are consistent with
an O6V--III spectral type. The primary's spectrum contains little or
no evidence of \ion{Si}{4} $\lambda\lambda$4089, 4116~\AA, \ion{Mg}{2}
$\lambda$4481~\AA, or \ion{C}{3} $\lambda$4070~\AA\ indicating a
temperature class O6 or earlier.  The temperature-sensitive ratio,
\ion{He}{1}/\ion{He}{2}~$\lambda$4026:\ion{He}{2} $\lambda$4200~\AA\
is $\sim$1:1 and also indicates a temperature class of O6. It is
difficult to distinguish \ion{He}{1} $\lambda$4009~\AA\ from the
continuum noise, but there appears to be little or no evidence of
\ion{He}{1} $\lambda$4144~\AA, and \ion{He}{1} $\lambda$4471~\AA\ is
comparatively weak.  Because \ion{Si}{4} absorption is weak even in an
O6V--III and we did not include \ion{He}{2} $\lambda$4686~\AA~or
$\lambda$4542~\AA\ in our spectral coverage, the luminosity class can
only be constrained to lie between III and V \citep[interpolated
from][]{FM05}. The bottom spectrum in Figure~\ref{S3comp} is a
composite of two spectra from the \citet{WF90} digital atlas, HD37043
(O9III) and HD93130 (O6III). Intrinsic \ion{He}{1}, \ion{He}{2},
\ion{C}{3}, \ion{Si}{4}, \ion{Mg}{2}, and hydrogen absorption are
labeled. We found the best agreement with Schulte~3 when HD37043 and
HD93130 were combined with an equal component luminosity ratio,
indicating the primary may in fact have a luminosity class of IV
\citep[interpolated from][]{FM05}. 

There is a discrepancy between the theoretical mass ratio
($q=0.64\pm0.02$ for an O6III \& O9III or $q=0.72\pm0.02$ for an O6V
\& O9III) obtained from \citet{FM05} and the mass ratio of
$q=0.44\pm0.08$ found in this study. We attribute this discrepancy to
either the uncertainty in mid-to-early O star masses \citep{Massey05},
or the possibility of significant mass transfer between the components
based on the hydrogen and helium emission line profiles
\citep{Miller07,Falceta06,Sana01}. It should be noted that an
uncertainty in the spectral type of each component could also
partially explain the discrepancy. For instance, an O6III primary and
O9.5V secondary results in a mass ratio of $q=0.45\pm0.02$
\citep{FM05}. However, this combination produces a luminosity ratio in
disagreement with the implied luminosity ratio of the double-lined
spectra and the light curve of Schulte~3. An O5III primary and O9.5III
secondary would also replicate the computed mass ratio within the
error \citep[$q=0.51\pm0.01$;][]{FM05}, but also suffers from the same
luminosity ratio disagreement. A likely scenario that explains the
mass ratio discrepancy may involve a combination of two or more
factors. Further analysis of the light curve for Schulte~3 \citep[in
prep]{Karen} and additional spectroscopic coverage is necessary
however, to verify the presence of a mass transfer mechanism and
determine what type of eclipsing system this is.

\section{Period Estimates of three Multi-lined Systems}
\begin{figure}
\epsscale{1.0}
\plotone{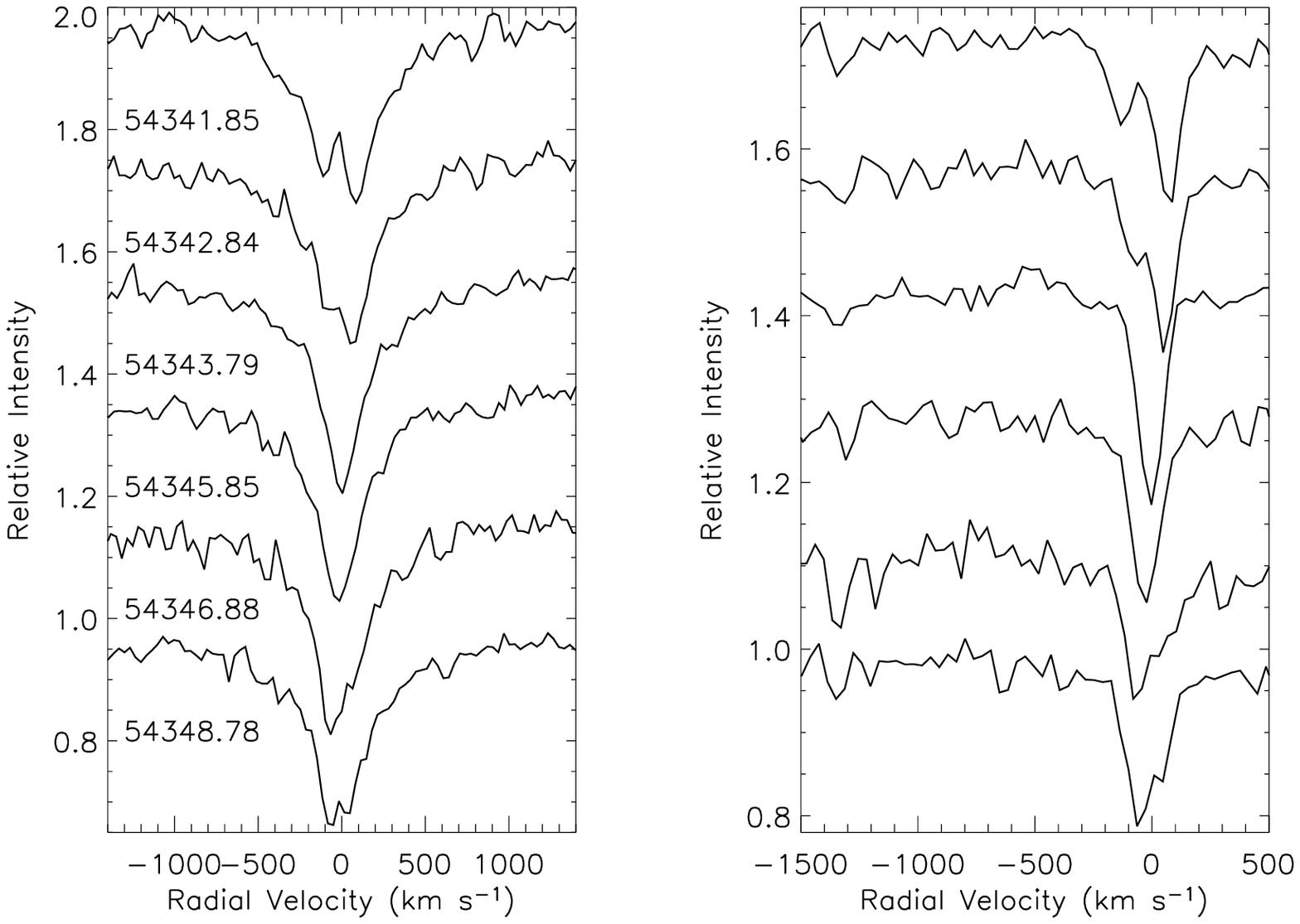}
\caption{H$\alpha$ $\lambda$6562.80~\AA\ and \ion{He}{1} $\lambda$5875.75~\AA\
absorption in velocity space for MT252 using same format as 
Figure~\ref{S3velprog}.
 \label{MT252pro}}
\end{figure}

\begin{figure}
\epsscale{1.0}
\plotone{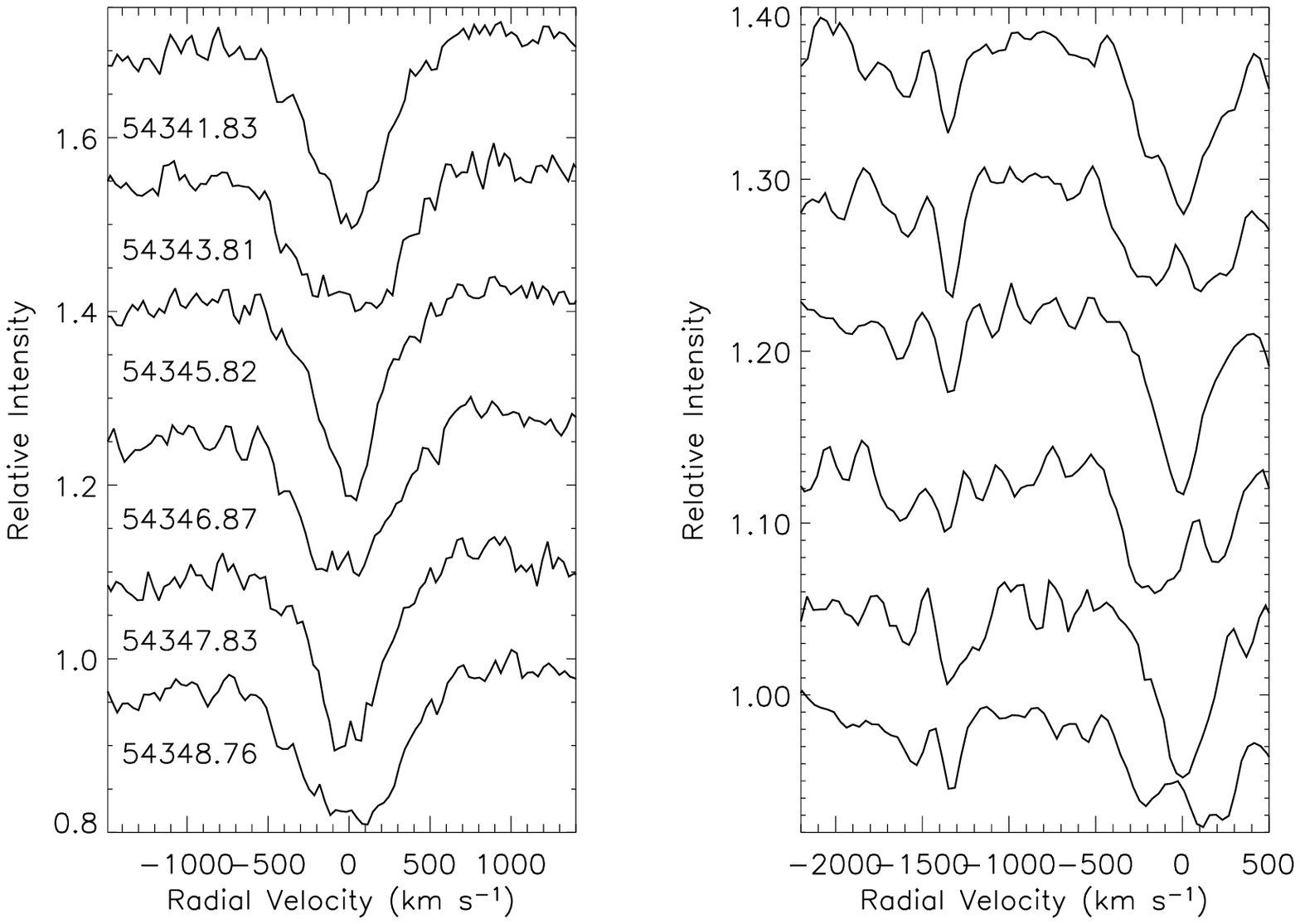}
\caption{H$\alpha$ $\lambda$6562.80~\AA\ and \ion{He}{1} $\lambda$5875.75~\AA\
absorption in velocity space for MT720 using same format as 
Figure~\ref{S3velprog}.
 \label{MT720pro}}
\end{figure}

\begin{figure}
\epsscale{1.0}
\plotone{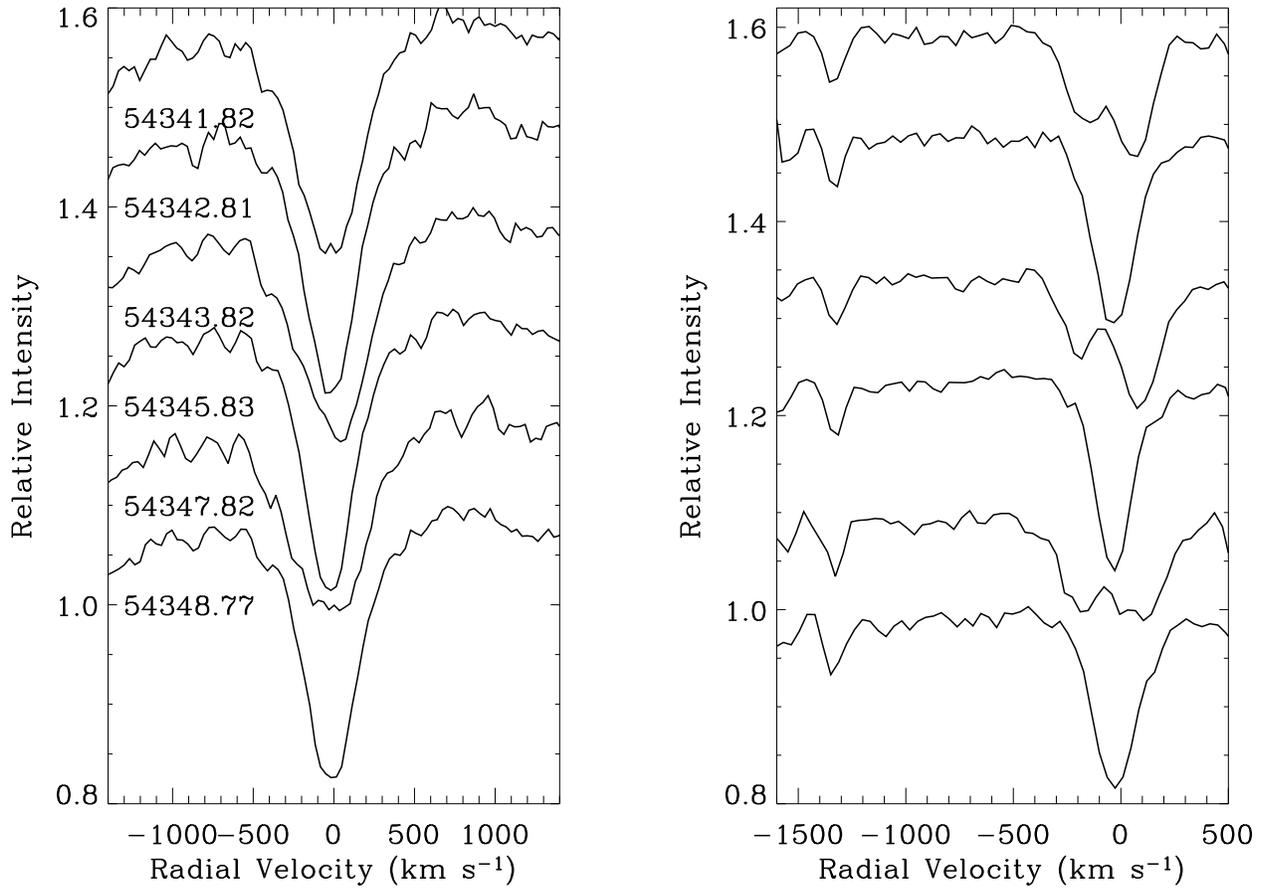}
\caption{H$\alpha$ $\lambda$6562.80~\AA\ and \ion{He}{1} $\lambda$5875.75~\AA\
absorption in velocity space for MT771 using same format as 
Figure~\ref{S3velprog}.
 \label{MT771pro}}
\end{figure}

The 2007 August/September dataset was not sufficient to fit solutions
to three double-lined systems but did allow for stricter constraints
on the orbital elements. We applied the same method used for Schulte~3
to measure radial velocities for the components of these
systems. Figures~\ref{MT252pro}, \ref{MT720pro}, \&\ \ref{MT771pro}
provide the time series for H$\alpha$ $\lambda6563$~\AA\ and
\ion{He}{1} $\lambda$5876~\AA\ in velocity space for MT252, MT720, and
MT771 (2007 August 28 through 2007 September 4). Each system was
introduced in \pap\ with approximate periods and secondary mass
estimates.

\subsection{The SB2 MT252}
For the first double-lined system, MT252, we obtained 23
observations from 1999--2007.  MT252 is classified as a B1.5III in
\pap. We revise the classification slightly in this study to
B2III. The primary shows weak \ion{C}{2}/\ion{O}{2} $\lambda$4070~\AA,
little or no \ion{Si}{4} $\lambda$4089~\AA, and a ratio of \ion{Si}{3}
$\lambda$4552:\ion{Si}{2} $\lambda\lambda$4128, 4430~\AA\ of over 2:1,
indicating a B2 temperature class.  The luminosity indicator,
\ion{Si}{3} 4552:\ion{He}{1} $\lambda$4387~\AA\ is nearly 2:1 and
points to a luminosity class of III. The secondary shows an absence of
\ion{Si}{4} $\lambda$4089~\AA\ and \ion{C}{2}/\ion{O}{2}
$\lambda$4070~\AA, making it no earlier than B1. There is also no
evidence of \ion{C}{2} $\lambda$4267 \AA\ or \ion{Si}{2}
$\lambda\lambda$4128, 4130 \AA, indicating it is likely no later than
a B2.  A \ion{He}{1} $\lambda$4121:\ion{He}{1} $\lambda$4144~\AA\
ratio of nearly 1:1 and the absence of \ion{N}{3} $\lambda$4634~\AA\
and \ion{O}{2} $\lambda\lambda$4640, 4650~\AA\ absorption places the
luminosity at V. We therefore conclude that the secondary is most
likely a B1V.

Of the 23 observations obtained, only six are of high enough S/N to
estimate \ion{He}{1} line velocities. These velocities listed in
Table~\ref{OC2} are relative to the heliocentric frame of rest. The
largest line separation for MT252 is seen on 2007 August 28
(Figure~\ref{MT252pro}), and a near reversal of spectral features is
seen within the $\sim$8 days covered, indicating a likely period of
18--19 days. The maximum line separation in \ion{He}{1}
$\lambda$5876~\AA\ is $\sim$192~\kms, and the $V_r$ semi-amplitudes
are $K_2=107\pm13$~\kms\ and $K_1=85\pm13$~\kms\ relative to a
systemic velocity of $\gamma=-18\pm10$~\kms estimated from the 2007
August 30 and September 1 spectra (where the errors are approximated
as they were with Schulte~3) indicating a mass ratio of
$q\simeq0.8\pm0.2$, in agreement with the theoretical mass ratio of
$q=0.9$ \citep[interpolated from][]{drilling}. Based on a partial
$V_r$ vs. time curve of this system, the eccentricity appears
low. Therefore, for an assumed circular orbit, semi-amplitudes of
$K_1=85\pm13$~\kms\ and $K_2=107\pm13$~\kms\ on 2007 August 28, and a
period of $P\sim18.5\pm0.5$~days, the calculated masses are
$M_1$sin$^3\simeq8\pm1$~\msun\ and $M_2$sin$^3i\sim6\pm1$~\msun. These
values also imply a primary and secondary semi-major axes of
$a_1$sin~$i\simeq22\pm3$~\rsun\ and $a_2$sin~$i\simeq27\pm3$~\rsun. It
should be noted that these values reflect lower limits in the absence
of a full orbital solution.

\subsection{The SB2 MT720}
\begin{figure}
\epsscale{1.0}
\plotone{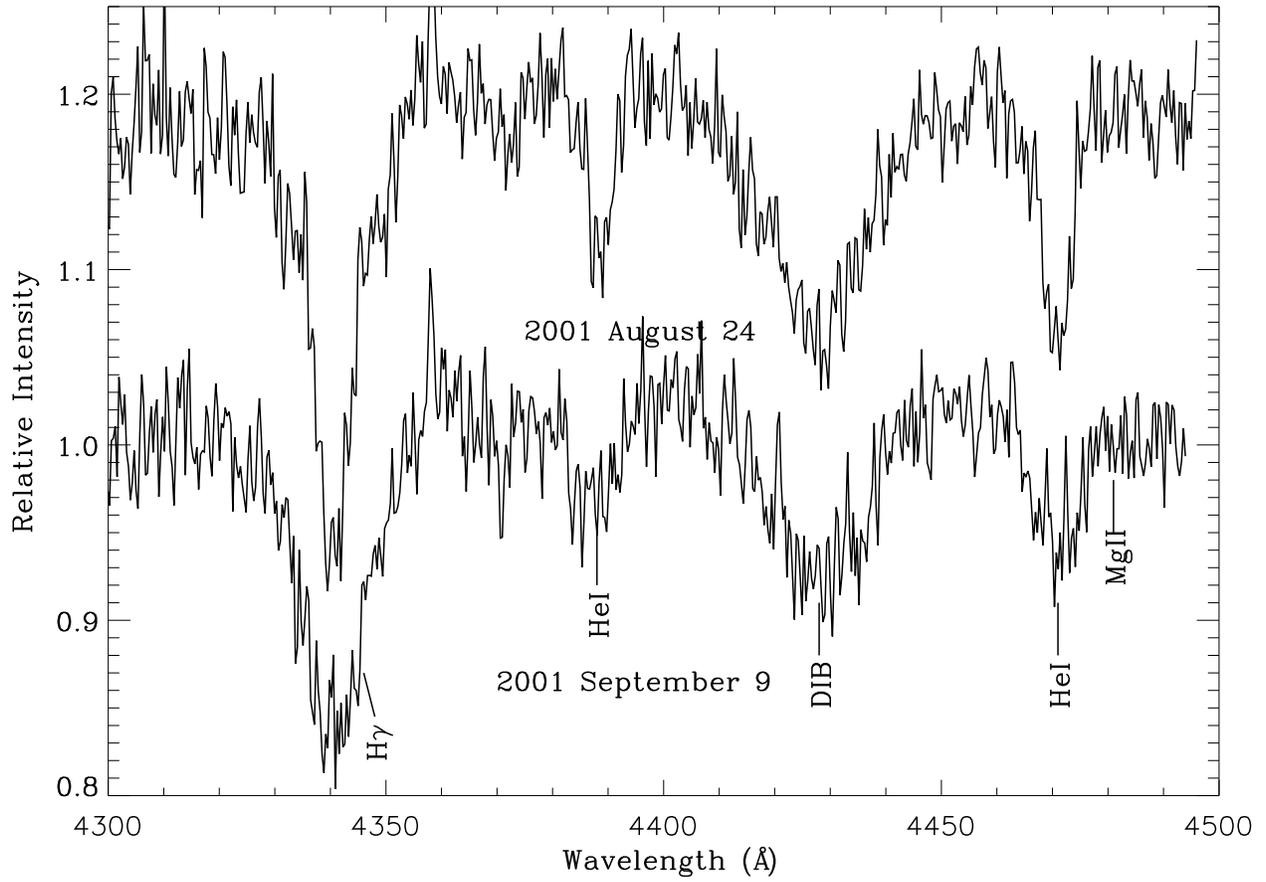}
\caption{A portion of the 2001 August 24 (top) and 2001 September~9 
(bottom) spectra for MT720, showing the nearly blended and unblended 
states respectively.
 \label{triple}}
\end{figure}

MT720 is composed of two components of apparently equal luminosity
based on the similar depth and ratios of hydrogen and \ion{He}{1}
lines, as illustrated in a small portion of the 2001 August 24 and
2001 September 9 observations in Figure~\ref{triple}. The top spectrum
(August 24) shows a nearly complete blend while the bottom spectrum
(September 9) shows the deblended state (with perhaps a slight
asymmetry in the red wing of the Balmer lines). The blended spectral
features seen on 2007 September 1 \&\ 3 (\textit{third and fifth
spectra down}) in Figure~\ref{MT720pro} indicate the systemic radial
velocity, $\gamma$, is near zero. While the quality of the data makes
it difficult to estimate an exact period, September 2 \&\ 4
(\textit{fourth and sixth spectra down}) indicate the period is
$\sim5$~days or less. The low S/N of the data makes it unclear whether
the spectra show only two components. Therefore we applied both a two
and three component model to both the hydrogen and \ion{He}{1} lines
in the attempt to deblend the spectral features. We obtained mixed
results with the two-Gaussian fit (fixed widths), which provided
velocity ratios varying from 0.4--0.9. Three Gaussians however,
provided only slightly better fits to the data and again yielded
varying velocity ratios. Therefore, we adopted the simpler two
component model. The absence of \ion{He}{2} in any of the spectra
indicates that neither of the two components are earlier than about B0.
The strength of \ion{He}{1} coupled with the appearance of
multi-component \ion{C}{3} $\lambda$4070~\AA\ and weak or absent
\ion{Mg}{2} $\lambda$4481~\AA\ indicates that both are probably B2 or
earlier. We therefore conclude the components lie between B0--B2 with
uncertain luminosity classes.

\subsection{The SB2 MT771}
We obtained 36 observations (most low S/N) of the double system
MT771. The primary is an O7V as indicated by the \ion{He}{2}
$\lambda$4200:\ion{He}{1} $\lambda$4144~\AA\ and \ion{He}{2}
$\lambda$4200:\ion{He}{1} $\lambda$4026~\AA\ ratios of nearly 6:1 and
1:1 respectively. The ratio of \ion{He}{2} $\lambda$4542:\ion{He}{1}
$\lambda$4387~\AA\ of just over 1:1 gives the luminosity class. The
secondary is a probable O9V as best indicated by the temperature
sensitive ratio of \ion{He}{2} $\lambda$4200:\ion{He}{1}
$\lambda$4121~\AA\ ($\sim$1:1) and luminosity-sensitive ratio of
\ion{Si}{4} $\lambda$4116:\ion{He}{1} $\lambda$4121~\AA\ (just under
1:1).

Figure~\ref{MT771pro} shows blended features for H$\alpha$
$\lambda$6563~\AA\ and \ion{He}{1} $\lambda$5876~\AA\ on days 2, 5,
and 8, suggesting periods of 1.5, 3, or 6 days.  The similar He line
profiles on days 1 and 3 suggest that one full period has elapsed,
leaving the $\sim$1.5 day period as the only viable option.  The
August 30 (\textit{third spectrum down}) spectrum shows a $V_r$
separation of 277 \kms\ ($-153\pm14$ \kms\ and $124\pm13$ \kms\
relative to the systemic velocity of $\gamma=-36\pm10.0$~\kms
estimated from the 2007 September 1 spectrum) and implies a mass ratio
of $q=0.8\pm0.1$, consistent with a theoretical mass ratio of
$q=0.68\pm0.02$ for an O7V ($25.9\pm0.6$ \msun) and O9V ($17.6\pm0.5$
\msun) \citep{FM05}. Assuming a circular orbit, period of $1.5$ days,
and minimum semi-amplitudes of $K_1=124\pm13$ \kms\ and $K_2=153\pm14$
\kms\ for the primary and secondary respectively, the minimum
semi-major axes are $a_1$sin~$i\sim2.6\pm0.3$~\rsun\ and
$a_2$sin~$i\sim3.2\pm0.3$~\rsun, and the masses are
$M_1$sin$^3i\sim1.8\pm0.2$~\msun\ and
$M_2$sin$^3i\sim1.5\pm0.2$~\msun. As with MT252, these values reflect
lower limits in the absence of a full orbital solution. No evidence
for interacting winds or mass transfer is seen. Therefore, using the
theoretical radii of $R_1=9.3\pm0.1$ and $R_2=7.6\pm0.1$ \citep{FM05},
this system would need to have an inclination of $i\lesssim20^\circ$
to avoid contact. This is consistent with the low inclination
necessary ($i\sim24$) to obtain the theoretical masses for an O7V and
O9V.

\section{Summary of Survey Results to Date}
\begin{figure}
\epsscale{1.0}
\plotone{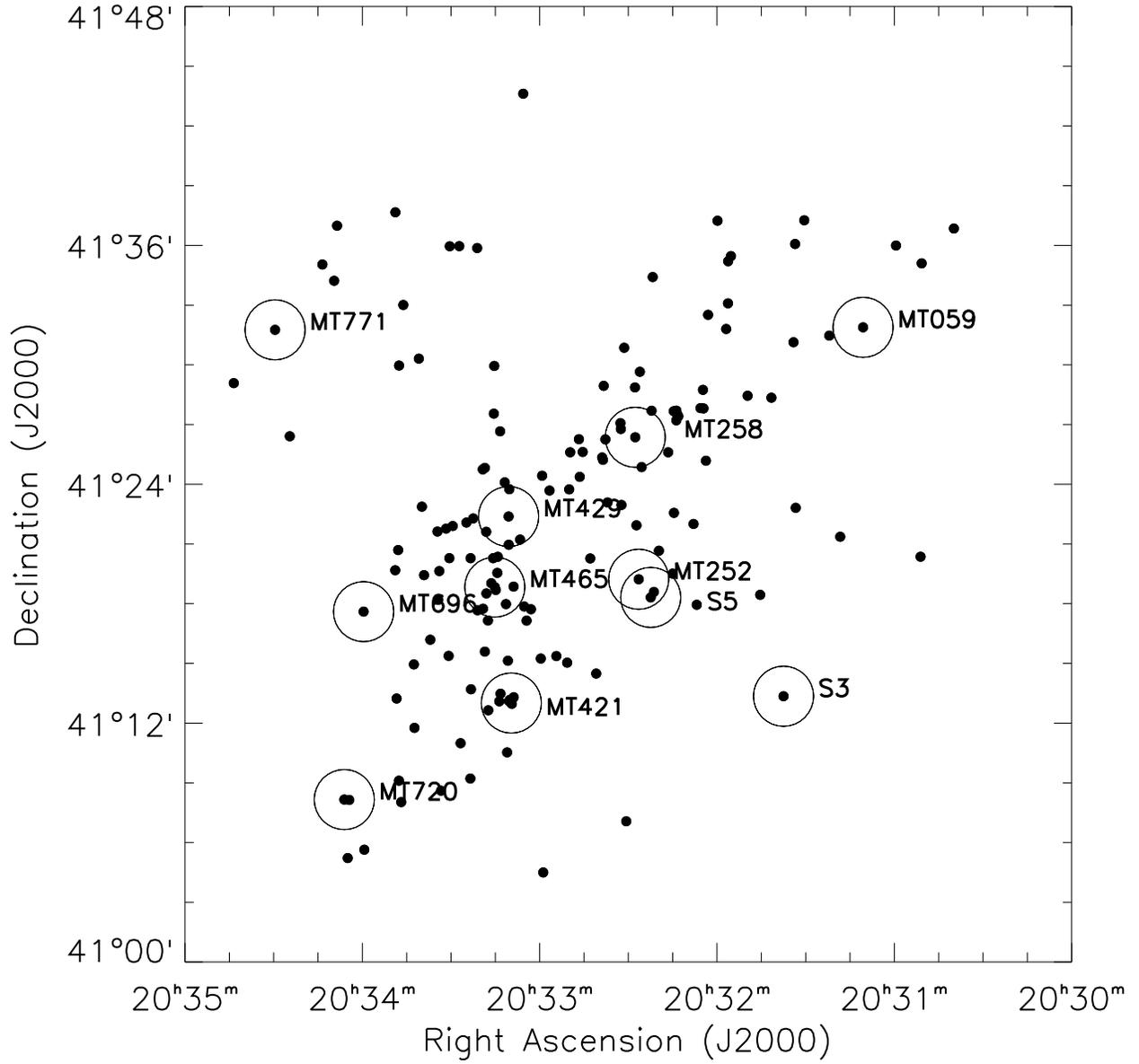}
\caption{A map of OB stars from this survey. Circles indicate the 
location of known OB binaries within this region of Cyg OB2.
 \label{bin_locat}}
\end{figure}

We have discovered the presence of six binary systems in the Cyg OB2
association as part of an ongoing survey to determine the distribution
of binary orbital parameters.  We presented the orbital solutions to
three new systems (MT059, MT258, and Schulte~3) and constrained the
orbital parameters to the three additional double-lined binaries
(MT252, MT720 \&\ MT771). Mass ratios for the six systems in this
study range from $q\geq0.18\pm0.01$ for MT258 to $q\simeq0.8\pm0.2$
for MT252, and have periods ranging from $\sim$1.5--19~days. Systemic
velocities deviate by $\sim$10--20 \kms\ from the cluster mean of
$-10.3$ \kms\ obtained in \pap, but they are still well within the
Association's radial velocity dispersion. These six systems more than
double the number of known binaries in this cluster and bring the
total number of uncovered OB binary/multiple systems in Cyg OB2 to
11. The locations of all 11 systems are shown in
Figure~\ref{bin_locat}. No evidence of grouping is
apparent. Table~\ref{Binaries} lists several key parameters for each
system, including the star designation, photometric and/or
spectroscopic binary type (\textit{Type}), spectral classifications
(\textit{S.C.}), period (\textit{P}), mass ratio when available
(\textit{q}), and literature references (\textit{Ref}).  Periods for
all 11 systems are relatively short ($<22$ days) and the corresponding
eccentricities are relatively small (0.03--0.11). With the exception of
MT421, all systems have early companions. However, the apparent
secondary mass distribution may depend on the sensitivity of the
associated surveys. 

Although our current sample encompasses only a small fraction of
systems in Cyg OB2, we can test the limited mass ratio statistics
against the hypothesis of \citet{PinStanek06}, that 45\%\ of massive
stars exist in ``twin'' systems with mass ratios of $q>0.95$. Of the
seven systems with mass ratio measurements in Table~\ref{Binaries}, we
find at most one system that may be considered a possible ``twin''
(MT252 with $q=0.8\pm0.2$). If this hypothesis holds true, binomial
statistics indicate the present distribution is likely only 8.7\% of
the time. On the other hand, if we consider $q<0.95$ for MT252, the
present distribution is likely only 1.5\% of the time. Under the
\citet{PinStanek06} hypothesis, we would expect to observe a
distribution of two to four ``twins'' in a sample of seven objects
(21.4\%, 29.2\%, and 23.9\% respectively). Though no definite
conclusions can yet be drawn about the agreement of these findings
with \citet{PinStanek06} owing to the limited number of systems
available, at least in the case of Cyg OB2, a ``twin'' heavy binary
distribution does not seem likely.

\acknowledgments

We thank the time allocation committees of the Lick, Keck, WIYN, and
WIRO Observatories for granting us observing time and making this
project possible. This paper has benefited greatly from the detailed
review and suggestions of an anonymous referee. We are also grateful
for support from the National Science Foundation through the Research
Experiences for Undergraduates (REU) program grant AST-0353760 and
through grant AST-0307778, and the support of the Wyoming NASA Space
Grant Consortium. The work of M.~V.~McSwain is supported by the
National Science Foundation, under grant AST-0401460 and she is also
grateful for institutional support from Lehigh University. We also
would like to graciously thank Sarah Bird, Georgi Chunev, Megan
Bagley, Emily May, Christopher Rodgers, Karen Kinemuchi, Brian Uzpen,
and Carolynn Moore for their help observing at WIRO though any weather
condition.

\textit{Facilities:} \facility{WIRO ()}, \facility{WIYN ()},
\facility{Shane ()}, \facility{Keck:I ()}

\clearpage

\begin{deluxetable}{lllccc}
\tabletypesize{\scriptsize}
\tablecaption{Summary of Observing Runs \label{obs.tab}}
\tablewidth{0pt}
\tablehead{
\colhead{Date} &
\colhead{Observatory} &
\colhead{Spectral} & 
\colhead{Grating} &
\colhead{Mean Spectral} & 
\colhead{Date} \\
\colhead{} &
\colhead{/Instrument} &
\colhead{Coverage} &
\colhead{} &
\colhead{Resolution} &
\colhead{Coverage} \\
\colhead{} &
\colhead{} &
\colhead{(\AA)} &
\colhead{(l mm$^{-1}$)} &
\colhead{(\AA)} &
\colhead{(HJD)}}
   
\startdata
1999 Jul 4--5       & Keck/HIRES      & 3890--6270 in 35 orders & 31.6   & 0.1 & 2,451,363--2,451,364 \\
1999 Jul 21--23     & Lick/Hamilton   & 3650--7675 in 81 orders & 31.6   & 0.1 & 2,451,381--2,451,383 \\
1999 Aug 21--23     & Lick/Hamilton   & 3650--7675 in 81 orders & 31.6   & 0.1 & 2,451,411--2,451,413 \\
1999 Oct 14--15     & Keck/HIRES      & 3700--5250 in 29 orders & 31.6   & 0.1 & 2,451,466--2,451,467 \\
2000 Jul 10--11     & Lick/Hamilton   & 3650--7675 in 81 orders & 31.6   & 0.1 & 2,451,736--2,451,737 \\
2000 Sep 18--19     & Keck/HIRES      & 3700--5250 in 29 orders & 31.6   & 0.1 & 2,451,805--2,451,806 \\
2001 Aug 24         & WIYN/Hydra      & 3800--4490 in order 2   & 1200   & 0.9 & 2,452,146            \\
2001 Sep 8--9       & WIYN/Hydra      & 3800--4490 in order 2   & 1200   & 0.9 & 2,452,161--2,452,162 \\
2004 Nov 28--30     & WIYN/Hydra      & 3800--4490 in order 2   & 1200   & 0.9 & 2,453,338--2,453,340 \\
2005 Jul 18--21     & WIRO/WIRO-Spec  & 3800--4490 in order 1   & 2400   & 2.5 & 2,453,570--2,453,573 \\
2005 Jul 18--20,22  & WIRO/WIRO-Spec  & 3800--4490 in order 1   & 2400   & 2.5 & 2,453,632--2,453,635 \\
2005 Oct 13         & WIRO/Longslit   & 4050--6050 in order 2   & 600    & 2.5 & 2,453,657 \\
2006 Jun 16--20     & WIRO/Longslit   & 3900--5900 in order 2   & 600    & 2.5 & 2,453,903--2,453,907 \\
2006 Jul 15--16,20  & WIRO/Longslit   & 3900--5900 in order 2   & 600    & 2.5 & 2,453,932--2,453,935 \\
2006 Sep 8--11      & WIYN/Hydra      & 3800--4490 in order 2   & 1200   & 0.9 & 2,453,987--2,453,990 \\
2006 Oct 7          & WIRO/Longslit   & 3900--5900 in order 2   & 600    & 2.5 & 2,454,016 \\
2007 Jun 28--30     & WIRO/Longslit   & 3900--5900 in order 2   & 600    & 2.5 & 2,454,280--2,454,282 \\
2007 Jul 4--6       & WIYN/Hydra      & 3820--4510 in order 2   & 1200   & 0.9 & 2,454,285--2,454,287 \\
2007 Aug 28--Sep 4  & WIRO/Longslit   & 5550--6850 in order 1   & 1800   & 1.5 & 2,454,341--2,454,348 
\enddata
\end{deluxetable}

\clearpage

\pagestyle{plaintop}

\begin{deluxetable}{
lrrrrr}
\tabletypesize{\tiny}
\tabletypesize{\scriptsize}
\tablewidth{0pt}
\tablecaption{Orbital Elements \label{orbparms.tab}}
\tablehead{
\colhead{Element} &
\colhead{MT059} & 
\colhead{MT258} & 
\colhead{Schulte 3} & 
\colhead{MT252} &
\colhead{MT771}}  
\startdata
$P$ (Days)            & 4.8527 (0.0002)             & 14.660 (0.002)    	   & 4.7464 (0.0002) & 18--19         & 1.5:          \\
$e$                   & 0.11 (0.04)                 & 0.03  (0.05)      	   & 0.070 (0.009)   & \nodata        & \nodata       \\
$\omega$ (deg)        & 212 (17)                    & 68 (71)           	   & 5.5 (0.7)       & \nodata        & \nodata       \\
$\gamma$ (\kms)       & -24.2 (1.7)                 & -20.8 (1.2)       	   & -26.4 (1.7)     &  -18 (10)      & -36 (10)      \\
$T_0$ (HJD-2,400,000) & 53916.72 (0.22)             & 53922.03 (2.87)   	   & 53987.81 (0.01) & 54341.9 (1.0)  & 54343.8 (0.5) \\
$K_{1}$ (\kms)        & 72.3 (2.3)                  & 41.2 (1.7)        	   & 113.2 (14.5)    & 85 (13)        & 124 (13)      \\
$K_{2}$ (\kms)        & \nodata                     & \nodata           	   & 256.7 (2.4)     & 107 (13)       & 153 (14)      \\
$M_{1}$ (\msun)       & 21.4 (0.6)\tablenotemark{a} & 21.4 (0.6)\tablenotemark{a}  & $>$ 17.2 (0.3)  & $>$ $8\pm1$    & $>$ 1.8 (0.2) \\
$M_{2}$ (\msun)       & 5.1 (0.3)--13.8 (0.6)       & 4.1 (0.2)--17.5 (1.0)        & $>$ 7.6 (1.4)   & $>$ $6\pm1$    & $>$ 1.5 (0.2) \\
$f(m)_1$ (\msun)      & 0.187 (0.018)               & 0.106 (0.013)     	   & 0.709 (0.273)   & \nodata        & \nodata       \\
$f(m)_2$ (\msun)      & \nodata                     & \nodata         	           & 8.289 (0.237)   & \nodata        & \nodata       \\
S.~C.$_1$             & O8V                         & O8V               	   & O6IV:           & B1.5III        & O7V           \\
S.~C.$_2$             & B                           & B                            & O9III           & B1V            & O9V           \\
$a_{1}$sin~$i$ (\rsun) & 6.9 (0.2)                   & 11.9 (0.5)        	   & 7.4 (0.9)       & 22 (3)         & 2.6 (0.3)     \\
$a_{2}$sin~$i$ (\rsun) & \nodata                     & \nodata           	   & 16.7 (0.2)      & 27 (3)         & 3.2 (0.3)     \\
$rms_1$ (\kms)        & 11.0                        & 6.7               	   & 38.7            & \nodata        & \nodata       \\
$rms_2$ (\kms)        & \nodata                     & \nodata              	   & 9.0             & \nodata        & \nodata       \\

\enddata
\tablecomments{Calculated errors are located in parentheses.}
\tablenotetext{a}{Theoretical masses adopted from \citet{FM05}.}

\end{deluxetable}

\clearpage

\pagestyle{plaintop}

\begin{deluxetable}{cccrrr}
\tabletypesize{\tiny}
\tabletypesize{\scriptsize}
\tablewidth{0pc}
\tablecaption{Ephemerides for MT059 \&\ MT258 \label{OC}}
\tablehead{
\colhead{Star} & 
\colhead{Date} & 
\colhead{$\phi$} & 
\colhead{$V_r$} &
\colhead{1$\sigma$ err} &
\colhead{$O-C$} \\
\colhead{} & 
\colhead{(HJD-2,400,000)} & 
\colhead{} &
\colhead{(\kms)} &
\colhead{(\kms)} &
\colhead{(\kms)}}
\startdata  
MT059...  &    51381.50 & 0.562  &      10.0 &     25.6 &      -5.8 \\
       &    51411.50 & 0.744  &        -89.1 &     31.6 &     -37.2 \\
       &    51413.50 & 0.156  &         -4.0 &     15.4 &      15.4 \\
       &    51467.91 & 0.369  &         51.8 &     13.4 &      11.2 \\
       &    51736.50 & 0.717  &        -32.4 &     32.8 &       8.2 \\
       &    51737.50 & 0.924  &        -98.0 &     14.6 &       5.2 \\
       &    51805.75 & 0.988  &       -106.7 &      6.0 &     -10.5 \\
       &    52146.70 & 0.248  &         15.7 &      4.1 &      -2.7 \\
       &    52162.66 & 0.538  &         26.9 &      4.4 &       4.7 \\
       &    53338.62 & 0.871  &        -90.3 &      4.7 &       6.1 \\
       &    53570.50 & 0.654  &        -26.3 &      5.2 &     -11.3 \\
       &    53571.50 & 0.860  &       -103.3 &      5.2 &      -9.4 \\
       &    53573.50 & 0.272  &         21.3 &      5.1 &      -4.3 \\
       &    53633.50 & 0.637  &        -15.6 &      4.9 &      -7.1 \\
       &    53634.50 & 0.843  &        -88.6 &      5.0 &       0.4 \\
       &    53636.50 & 0.255  &         26.4 &      5.6 &       5.6 \\
       &    53657.50 & 0.582  &         34.7 &      4.3 &      24.7 \\
       &    53903.72 & 0.322  &         28.3 &     15.6 &      -7.6 \\
       &    53903.73 & 0.324  &         24.3 &     11.1 &     -11.9 \\
       &    53903.74 & 0.325  &         32.6 &     11.0 &      -3.7 \\
       &    53903.84 & 0.346  &         37.0 &     11.5 &      -1.8 \\
       &    53903.84 & 0.347  &         39.1 &     12.0 &       0.0 \\
       &    53904.70 & 0.523  &          5.7 &     11.6 &     -20.0 \\
       &    53904.71 & 0.525  &         35.0 &     11.5 &       9.6 \\
       &    53904.86 & 0.558  &          7.0 &     11.1 &     -10.1 \\
       &    53904.89 & 0.563  &         22.9 &     11.0 &       7.1 \\
       &    53905.74 & 0.739  &        -82.5 &     13.8 &     -33.0 \\
       &    53905.82 & 0.755  &        -53.0 &     11.0 &       3.4 \\
       &    53905.84 & 0.758  &        -48.6 &     10.9 &       9.1 \\
       &    53906.72 & 0.939  &        -99.7 &     12.1 &       3.4 \\
       &    53906.72 & 0.941  &       -100.0 &     11.4 &       3.1 \\
       &    53906.87 & 0.972  &       -112.8 &     11.7 &     -13.1 \\
       &    53907.77 & 0.156  &        -22.4 &     11.1 &      -2.7 \\
       &    53907.86 & 0.175  &        -10.3 &     11.0 &       0.0 \\
       &    53932.73 & 0.301  &         42.1 &     11.0 &       9.9 \\
       &    53932.83 & 0.321  &         77.8 &     13.3 &      42.0 \\
       &    53932.84 & 0.322  &         67.8 &     12.4 &      31.8 \\
       &    53935.77 & 0.927  &        -91.2 &     13.2 &      12.1 \\
       &    53935.85 & 0.943  &        -71.5 &     12.0 &      31.4 \\
       &    53987.70 & 0.628  &         -1.2 &      8.3 &       3.8 \\
       &    53988.74 & 0.843  &        -79.0 &      6.3 &      10.0 \\
       &    53989.66 & 0.031  &        -82.7 &      4.2 &      -0.9 \\
       &    53990.85 & 0.277  &         30.2 &      5.1 &       3.2 \\
       &    54285.93 & 0.084  &        -48.9 &      6.4 &       8.0 \\
MT258...  &    51467.91 & 0.262 &  -11.4 &    7.7 &    1.3  \\
       &    51805.74 &   0.223 &     1.4 &    4.8 &    3.5  \\
       &    52146.70 &   0.386 &    21.7 &    3.9 &    6.0  \\
       &    52161.81 &   0.368 &    12.2 &    3.8 &    1.3  \\
       &    52162.66 &   0.424 &    -1.7 &    3.8 &    0.1  \\
       &    53338.62 &   0.867 &   -44.4 &    3.6 &    9.7  \\
       &    53340.59 &   0.995 &   -57.4 &    4.0 &    3.0  \\
       &    53571.50 &   0.005 &   -36.3 &    4.4 &   -7.1  \\
       &    53573.50 &   0.135 &   -57.2 &    4.5 &   -0.1  \\
       &    53632.50 &   0.970 &   -62.0 &    6.3 &   -2.4  \\
       &    53633.50 &   0.035 &   -72.8 &    4.7 &  -11.5  \\
       &    53903.76 &   0.603 &    12.5 &    11.1 &   -4.8   \\
       &    53903.77 &   0.604 &    13.3 &    11.3 &   -4.1   \\
       &    53904.76 &   0.668 &    11.9 &    11.0 &   -9.0   \\
       &    53904.77 &   0.668 &    10.3 &    11.6 &  -10.5   \\
       &    53906.75 &   0.797 &    -4.1 &    11.0 &   -9.0   \\
       &    53906.76 &   0.798 &     1.7 &    11.2 &   -3.1   \\
       &    53906.76 &   0.798 &    -5.6 &    11.1 &  -10.3   \\
       &    53907.81 &   0.867 &   -17.7 &    10.7 &   -4.5   \\
       &    53907.82 &   0.867 &   -12.6 &    10.8 &    0.6   \\
       &    53932.75 &   0.488 &    11.3 &    10.9 &   -3.3   \\
       &    53932.86 &   0.495 &    34.9 &    11.1 &   19.2   \\
       &    53935.81 &   0.686 &     5.3 &    12.6 &   -3.3   \\
       &    53935.86 &   0.690 &    19.6 &    11.5 &   11.6   \\
       &    53987.70 &   0.060 &   -38.5 &    4.6 &    1.5  \\
       &    53988.74 &   0.127 &   -17.0 &    4.4 &    6.5  \\
       &    53989.66 &   0.187 &   -13.2 &    3.7 &   -4.7  \\
       &    53989.78 &   0.195 &   -11.3 &    4.2 &   -4.7  \\
       &    53990.85 &   0.264 &     8.2 &    4.3 &   -0.6  \\
       &    54285.93 &   0.446 &    19.3 &    4.0 &   -1.6  \\
       &    54286.66 &   0.493 &    21.2 &    5.7 &    2.7  \\
       &    54342.88 &   0.708 &    10.9 &   13.6 &   -0.1  \\        
       &    54343.76 &   0.768 &    23.9 &   14.0 &    5.1  \\
       &    54345.82 &   0.908 &    22.0 &   13.8 &    7.7  \\
       &    54347.81 &   0.044 &    -6.7 &   13.6 &   10.0  \\
\enddata 

\end{deluxetable}

\clearpage
\pagestyle{plaintop}

\begin{deluxetable}{cccrrrr}
\tabletypesize{\tiny}
\tabletypesize{\scriptsize}
\tablewidth{0pc}
\tablecaption{Ephemerides for Schulte~3, MT252, \&\ MT771 \label{OC2}}
\tablehead{
\colhead{Star} & 
\colhead{Date} & 
\colhead{$\phi$} & 
\colhead{$V_{r1}$} &
\colhead{$O_1-C_1$} &
\colhead{$V_{r2}$} &
\colhead{$O_2-C_2$} \\ 
\colhead{} & 
\colhead{(HJD-2,400,000)} &
\colhead{} &
\colhead{(\kms)} &
\colhead{(\kms)} &
\colhead{(\kms)} &
\colhead{(\kms)}}
\startdata  
Schulte~3..... &  53987.69 & 0.974   &   -111 (20) &  35.4   &   250 (8) &   2.9    \\
               &  53988.74 & 0.195   &     12 (20) &  63.7   &    13 (8) &  -7.7    \\
     	       &  53989.65 & 0.387   &     78 (20) &  17.0   &  -224 (8) &   6.1    \\
     	       &  53989.77 & 0.412   &     46 (20) & -22.6   &  -244 (8) &   2.2    \\
               &  53990.85 & 0.639   &     90 (20) &  47.9   &  -179 (8) &  -5.1    \\
     	       &  54285.93 & 0.809   &   -123 (20) & -56.1   &    91 (8) &  15.8    \\
     	       &  54341.80 & 0.580   &    112 (20) &  46.7   &  -229 (8) &   0.3    \\
     	       &  54342.72 & 0.773   &    -10 (20) &  31.6   &    12 (8) &  -6.6    \\
     	       &  54343.83 & 0.007   &   -139 (20) &   7.7   &   243 (8) &  -2.2    \\
     	       &  54344.76 & 0.203   &    -20 (20) &  25.5   &    13 (8) &   6.5    \\
     	       &  54345.78 & 0.418   &     73 (20) &   2.7   &  -257 (8) &  -7.4    \\
     	       &  54346.79 & 0.631   &     69 (20) &  23.1   &  -179 (8) &   3.9    \\
     	       &  54347.75 & 0.833   &    -41 (20) &  42.9   &   103 (8) & -10.3    \\
     	       &  54348.84 & 0.063   &   -125 (20) &   8.0   &   211 (8) &   0.9    \\
MT252.....     &  54341.85 & \nodata &  67 (8)     & \nodata & -125 (8)  & \nodata \\
               &  54342.84 & \nodata &  1:         & \nodata &  -83:     & \nodata \\
               &  54343.79 & \nodata & -9:         & \nodata & \nodata   & \nodata \\
               &  54345.85 & \nodata & -27:        & \nodata & \nodata   & \nodata \\
               &  54346.88 & \nodata & -63:        & \nodata & \nodata   & \nodata \\
               &  54348.78 & \nodata & -64:        & \nodata &   41:     & \nodata \\
MT771.....     &  54341.81 & \nodata & 63 (10)     & \nodata & -165 (10) & \nodata  \\
               &  54342.81 & \nodata & \nodata     & \nodata & -11:      & \nodata  \\
               &  54343.82 & \nodata & 88 (8)     & \nodata & -189 (10) & \nodata  \\
               &  54345.83 & \nodata & -36 (10)    & \nodata & -36 (10)  & \nodata  \\
               &  54347.82 & \nodata & \nodata     & \nodata &  89:      & \nodata  \\
               &  54348.77 & \nodata & \nodata     & \nodata & -26:      & \nodata  \\
\enddata

\end{deluxetable}

\clearpage

\pagestyle{plaintop}

\begin{deluxetable}{lccccl}
\tabletypesize{\tiny}
\tabletypesize{\scriptsize}
\tablewidth{0pc}
\tablecaption{OB Binaries in Cyg OB2 \label{Binaries}}
\tablehead{
\colhead{Star} & 
\colhead{Type} & 
\colhead{S.C.} & 
\colhead{P} &
\colhead{q} &
\colhead{Ref.} \\
\colhead{} &
\colhead{} &
\colhead{} &
\colhead{(days)} &
\colhead{} &
\colhead{}}
\startdata
MT059      & SB1     & O8V \&\ B                          & 4.8527 (0.0002)      & 0.24 (0.02)--0.64 (0.03)  & 1 \\
MT252      & SB2     & B2III \&\ B1V                      & 18--19               & 0.8 (0.02)                & 1 \\
MT258      & SB1     & O8V \&\ B                          & 14.660 (0.002)       & 0.19 (0.01)--0.84 (0.05)  & 1 \\
MT421      & EA      & O9V \&\ B9V--A0V                   & 4.161   	         & \nodata                   & 2 \\
MT429      & EA      & B0V \&\ ??                         & 2.9788  	         & \nodata                   & 2 \\
MT696      & EW/KE   & O9.5V \&\ early B                  & 1.46    	         & \nodata                   & 3 \\
MT720      & SB2     & early B \&\ early B                & $<$ 5                & \nodata                   & 1 \\
MT771      & SB2     & O7V \&\ O9V                        & 1.5:                 & 0.8 (0.1)	             & 1 \\
Schulte 3  & SB2/EA: & O6IV: \&\ O9III                    & 4.7464 (0.0002)      & 0.44 (0.08)               & 1,4 \\
Schulte 5  & EB      & O7Ianfp \&\ Ofpe/WN9 (\&\ B0V?)    & 6.6                  & 0.28 (0.02)               & 5,6,7,8,9,10 \\	
Schulte 8a & SB2     & O5.5I \&\ O6:                      & 21.908 	         & 0.86 (0.04)               & 11,12 \\
\enddata

\tablecomments{Photometric types EW/KE, EA, and EB stand for Contact
system of the W UMa type (ellipsoidal; $P<1$ day), Algol type (near
spherical), and $\beta$ Lyr type (ellipsoidal; $P>1$ day)
respectively.}

\tablerefs{
(1) This study; 
(2) \citet{PJ98}; 
(3) \citet{Rios04};
(4) \citet[in prep]{Karen}; 
(5) \citet{Wilson48}; 
(6) \citet{Wilson51}; 
(7) \citet{Mics53};
(8) \citet{Wal73}; 
(9) \citet{Contreras97}; 
(10) \citet{Rauw99}; 
(11) \citet{Romano69}; 
(12) \citet{Debeck04}}

\end{deluxetable}

\clearpage

\clearpage

\clearpage

\clearpage

\clearpage

\clearpage

\clearpage

\clearpage

\clearpage

\clearpage

\begin{thebibliography}{}
\bibitem[Bonnell et al.(1998)]{Bonnell98}
Bonnell, I. A., Bate, M. R., \&\ Zinnecker, H. 1998, MNRAS, 198, 93
\bibitem[Contreras et al.(1997)]{Contreras97}
Contreras, M.~E., Rodriguez, L.~F., Tapia, M., Cardini, D., Emanuele, A., 
Badiali, M., \&\ Persi, P. 1997, ApJ, 488, 153
\bibitem[De Becker et al.(2004)De Becker, Rauw, and Manfroid]{Debeck04}
De Becker, M., Rauw, G., \&\ Manfroid, J. 2004, A\&A, 424, L39
\bibitem[Drilling \& Landolt(2000)]{drilling}
Drilling, J. S., \& Landolt, A. U. 2000, in Astrophysical Quantities, 
ed. A.~N. Cox (4th ed.; New York; Springer), 381
\bibitem[Evans et al.(2006)]{Evans06}
Evans, C. J., Lennon, D. J., Smartt, S. J., \&\ Trundle C. 
2007, A\&A, 464, 289
\bibitem[Falceta-Gon\c{c}alvez et al.(2006)]{Falceta06}
Falceta-Gon\c{c}alvez, D., Abraham, Z., Jatenco-Pereira, V. 2006, MNRAS, 371, 1295
\bibitem[Fullerton et al.(1996)]{Fullerton96}
Fullerton, A.~W.,Gies, D.~R. , \&\ Bolton C.~T. 1996, ApJS, 103, 475 
\bibitem[Garmany et al.(1980)]{Garmany80}
Garmany, C.~D., Conti, P.~S., \&\ Massey, P. 1980, ApJ, 242, 1063
\bibitem[Gies(1987)]{Gies87}
Gies, D.~R., 1987, ApJS, 64, 545
\bibitem[Hanson (2003)]{Hanson03}
Hanson, M.~M. 2003, ApJ, 597, 957
\bibitem[Harries et al.(2002)]{Harries02}
Harries, T.~J., Howarth, I.~D., \&\ Evans, C. 2002, MNRAS, 337, 341
\bibitem[Hillwig et al.(2006)]{Hillwig06}
Hillwig, T. C., Gies, D. R., Bagnuolo, W. G., Jr., Huang, W., 
McSwain, M. V., \&\ Wingert, D. W. 2006, ApJ, 639, 1069
\bibitem[Kiminki et al.(2007)]{Kiminki07}
Kiminki, D. C., et al. 2007, ApJ, 664, 1120
\bibitem[Kinemuchi et al.(2008)]{Karen}
Kinemuchi et al. 2008, in prep 
\bibitem[Kobulnicky \&\ Fryer(2007)]{Chip07}
Kobulnicky, H. A., \&\ Fryer, C. L. 2007, ApJ, 670, 747
\bibitem[Kouwenhoven et al.(2007)]{Kouwenhoven07}
Kouwenhoven, M.~B.~N., Brown, A.~G.~A., Portegies~Zwart, S.~F., \&\ 
Kaper, L. 2007, A\&A, 474, 77
\bibitem[Krumholz(2005)]{Krum05} 
Krumholz, M. R., 2005, ASPC, 352, 31
\bibitem[Kurtz et al.(1991)] {xcsao} 
Kurtz, M. J., Mink, D. J.,
Wyatt, W. F., Fabricant, D. G., Torres, G., Kriss, G. A., \&\ Tonry,
J. L. 1991, in Astronomical Data Analysis Software and Systems I, ASP
Conf. Ser., Vol. 25, eds. D.M. Worrall, C. Biemesderfer, and
J. Barnes, p. 432
\bibitem[Lanz \&\ Hubeny(2003)]{LHub2003}
Lanz, T., \&\ Hubeny, I. 2003, ApJS, 146, 417 
\bibitem[Larson(2001)]{Larson01}
Larson, R. B. 2001, IAU Symposium, 200, 93
\bibitem[Martins et al.(2005)Martins, Schaerer, and Hillier]{FM05} 
Martins, F., Schaerer, D., \&\ Hillier, D.~J. 2005, A\&A, 436, 1049
\bibitem[Massey \&\ Thompson(1991)]{MT91} 
Massey, P., \&\  Thompson, A.~B. 1991, AJ, 101, 1408
\bibitem[Massey et al.(2005)]{Massey05}
Massey, P., Puls, J., Pauldrach, A. W. A., Bresolin, F., Kudritzki, 
R. P., \&\ Simon, T. 2005, ApJ, 627, 477
\bibitem[McSwain(2003)]{mcswain03}
McSwain, M. V. 2003, ApJ, 595, 1124
\bibitem[Mermilliod(1995)]{Merm95}
Mermilliod, J. C. 1995, in Information \& On-Line Data in Astronomy,
 ed. D. Egret, \&\ M. A. Albrecht (Kluwer Academic Publishers, Dordrecht), 127
\bibitem[Miczaika (1953)]{Mics53}
Miczaika, G. R. 1953, PASP, 65, 141
\bibitem[Miller et al.(2007)]{Miller07}
Miller, B., Budaj, J., Richards, M., Koubsk\'y, P., \&\ Peters, 
G. 2007, ApJ, 656, 1075
\bibitem[Morbey \& Brosterhus(1974)]{Morbey74}
Morbey, C. L., \&\ Brosterhus, E. B., 1974, PASP, 86, 455
\bibitem[Pigulski \&\ Kolaczkowski(1998)Pigulski \&\ Kolaczkowski]{PJ98}
Pigulski, A., \&\ Kolaczkowski, Z. 1998, MNRAS, 298, 753
\bibitem[Pinsonneault \&\ Stanek (2006)]{PinStanek06}
Pinsonneault, M.~H., Stanek, K.~Z. 2006, ApJ, 639, 67
\bibitem[Rauw et al.(1999)]{Rauw99}
Rauw, G., Vreux, J.~M., \&\ Bohannan, B. 1999, ApJ 517, 416
\bibitem[Rios \&\ DeGioia-Eastwood(2004)]{Rios04}
Rios, L. Y., \&\ DeGioia-Eastwood, K. 2004, BAAS, 205, No. 09.05
\bibitem[Roberts et al.(1987)]{Roberts87}
Roberts, D.~H., Leh\'{a}r, J., \& Dreher, J. W., 1987, AJ, 93, 968 
\bibitem[Romano(1969)]{Romano69}
Romano, G. 1969, MmSAI, 40, 375
\bibitem[Sana et al.(2001)]{Sana01}
Sana, H., Rauw, G., Gosset, E. 2001, A\&A 370, 121
\bibitem[Schulte(1958)]{Schulte58}
Schulte, D.~H. 1958, AJ, 128, 41
\bibitem[Snow et al.(2002)Snow, Zukowski, and Massey]{Snow02}
Snow, T., Zukowski, D., \&\ Massey, P. 2002, ApJ, 578, 877
\bibitem[Walborn(1973)]{Wal73}
Walborn, N.~R. 1973, ApJ, 180, L35
\bibitem[Walborn \&\ Fitzpatrick(1990)]{WF90}
Walborn, N. R. \&\ Fitzpatrick, E. L. 1990, PASP, 102, 379 
\bibitem[Wilson (1948)]{Wilson48}
Wilson, O.~C. 1948, PASP, 60, 385
\bibitem[Wilson \&\ Abt (1951)]{Wilson51}
Wilson, O.~C., Abt, A. 1951, ApJ, 144, 477
\bibitem[Wolf et al.(2006)]{Wolf06}
Wolff, S. C., Strom, S. E., Dror, D., Lanz, L., \&\ Venn, K. 2006, 
AJ, 132, 749 
\end{thebibliography}
\end{document}